\documentclass{svjour3}                     
\usepackage{color}
\usepackage{amssymb,amsfonts}
\usepackage{graphicx}
\usepackage{latexsym}
\journalname{Journal of Modern Physics: http://www.scirp.org/journal/jmp}
\begin{document}
\large
\title{\bf A New Formulation of Quantum Mechanics}



\author{Arbab I. Arbab  \emph{and}   Faisal A. Yassein 
}


\institute{Arbab I. Arbab \at
              Department of Physics, Faculty of Science, University of Khartoum, P.O. Box 321, Khartoum
11115, Sudan \\
              Tel.: +249-122-174858\\
              Fax: +249-183780539\\
              \email{arbab.ibrahim@gmail.com, aiarbab@uofk.edu}           
           \and
           F. A. Yassein \at
             Department of Physics, Faculty of Science, Alneelain  University, P.O. Box 12702, Khartoum, Sudan\\
              \email{f.a.yassein@gmail.com}
}

\date{ }

\maketitle

\begin{abstract}
A new formulation of quantum mechanics based on differential commutator brackets is developed. We have found a wave equation representing the fermionic particle. In this  formalism, the continuity equation mixes the Klein-Gordon and Schrodinger probability density while keeping the Klein -Gordon and Schrodinger current unaltered. We have found time and space transformations under which Dirac's  equation is invariant. The invariance of Maxwell's equations under these transformations shows that the electric and magnetic fields of a moving charged particle are perpendicular to the  velocity of the propagating particle. This formulation agrees with the quaternionic formulation recently  developed  by Arbab.
\end{abstract}

\section{Introduction}
\baselineskip=20pt
Schrodinger's equation was used to explain and
describe all phenomena in atomic physics. However, after the development of the theory
of special relativity by Einstein, there was a need to unify quantum mechanics
and special relativity into a single \emph{Relativistic Quantum Theory}. Despite the success of
Schrodinger's equation in describing quite accurately the Hydrogen spectrum and giving
correct predictions for a large amount of spectral data, this equation is not invariant under
Lorenz transformations. In other words Schrodinger's equation is not relativistic and is
only an approximation valid at the non-relativistic limit when the velocities of the
particles involved are much smaller than the speed of light.

Quantum mechanics has been formulated by assigning an operator for any  dynamical  observable.  In Heisenberg formalism, the operator is governed by a commutator bracket. The fundamental commutator bracket relates to the position, and momentum is given by $[x\,, p_x]=i\hbar$. The commutator bracket generalizes the Poisson bracket of classical mechanics. If an operator commutes with Hamiltonian of the system, then the dynamical variable corresponding to that operator is said to be conserved.  An equation compatible with Lorentz transformation  guarantees its applicability to any inertial frame. Such an equation is symmetric in space-time. Thus a symmetric space-time formulation of any theory will generally guarantee the universality of the theory. However, Schrodinger equations does'nt exhibit this feature because it is not symmetric in space and time.  To remedy this problem, Klein and Gordon looked for an equation which is second order in space and time and consequently obtained the Klein-Gordon equation (KG). The probability density in this theory is found to be non-positive definite. Consequently, Dirac thought for a linear equation in space and time that has no such a problem. He obtained the familiar Dirac equation with a positive definite probability. However, the probability in KG formalism is later on (from a theoretic field point of view) interpreted as a charge density rather than a probability density which could be positive or negative [1].

With these motivation, we adopt a differential commutator bracket involving first order space and time derivative operators to formulate the Maxwell equations and quantum mechanics. This is in addition to our recent quaternionic formulation of physical laws, where we have shown that many physical equations are found to emerge from a unified form of physical variables [2]. Moreover, using quaternions, we have recently shown that quantum mechanics can be formulated in a set of three equations [5].  In such a  formulation,  the Dirac and Klein-Gordon  equations emerge from a set of three equations obtained from the application of an eigen-value problem of the linear momentum.

We aim in this paper to derive the equation of motion of the quantum system by applying the vanishing  differential commutator brackets. It is interesting to note that these commutator brackets are Lorentz invariant. Moreover, $\left[\triangle t\,, \triangle x\right]=\left[ \triangle t'\,, \triangle x'\right]$, where $t$ is time, $\vec{\nabla}=\frac{\partial}{\partial x}\hat{i}+ \frac{\partial}{\partial y} \hat{j}+\frac{\partial}{\partial z}\hat{k}$, $\Delta x=x_2-x_1$ and $t', x'$ are the moving time and space coordinates. We know that the second order partial derivatives commute for space-space variables. We don't assume here that this property is a priori for space and time. To guarantee this, we eliminate the time derivative of a quantity that is acted by a space ($\nabla$) derivative followed by a time derivative, and vice versa. In expanding the differential commutator brackets, we don't commute time and space derivative, but rather eliminate the time derivative by the space derivative, and vice versa. These linear differential commutator brackets may enlighten us to quantize these physical quantities.
By employing the differential commutator brackets of the vector $\vec{A}$ and scalar potential $\varphi$, we have derived Maxwell equations without invoking any  a priori physical law [3]. We would like here to  apply the differential commutator brackets to  explore quantum  mechanics.

\section{Differential commutators algebra}
Define the three linear differential commutator brackets as follows [3];
\begin{equation}\label{1}
    \left[\frac{\partial}{\partial t}\,, \vec{\nabla}\right]=0\,,\qquad    \left[\frac{\partial}{\partial t}\,, \vec{\nabla}\cdot\right]=0\,,
\qquad  \left[\frac{\partial}{\partial t}\,, \vec{\nabla}\times\right]=0\,,
\end{equation}
where $\vec{\nabla}$ and $\frac{\partial}{\partial t}$ are the space and time derivatives.
 \\
For a scalar $\varphi$ and a vector $\vec{A}$, one finds that
\begin{equation}\label{1}
    \left[\frac{\partial}{\partial t}\,, \vec{\nabla}\right]\varphi= \frac{\partial }{\partial t}\left(\vec{\nabla}\varphi\right)-\vec{\nabla}\left(\frac{\partial \varphi }{\partial t}\right)\,,
\end{equation}
and
\begin{equation}\label{1}
    \left[\frac{\partial}{\partial t}\,, \vec{\nabla}\cdot\right]\vec{A}= \frac{\partial }{\partial t}\left(\vec{\nabla}\cdot\vec{A}\right)-\vec{\nabla}\cdot\left(\frac{\partial \vec{A} }{\partial t}\right)\,,\qquad \left[\frac{\partial}{\partial t}\,, \vec{\nabla}\times\right]\vec{A}= \frac{\partial }{\partial t}\left(\vec{\nabla}\times\vec{A}\right)-\vec{\nabla}\times\left(\frac{\partial \vec{A} }{\partial t}\right)\,.
\end{equation}
Moreover, one can show that\footnote{See the Appendix for other identities.}
\begin{equation}\label{1}
    \left[\frac{\partial}{\partial t}\,, \vec{\nabla}\cdot\right](\varphi\vec{A})= \varphi\{\left[\frac{\partial}{\partial t}\,, \vec{\nabla}\cdot\right]\vec{A}\}+\{ \left[\frac{\partial}{\partial t}\,, \vec{\nabla}\right]\varphi\}\cdot\vec{A}\,,
\end{equation}

$$
    \left[\frac{\partial}{\partial t}\,, \vec{\nabla}\times\right](\varphi\vec{A})= \varphi\{\left[\frac{\partial}{\partial t}\,, \vec{\nabla}\times\right]\vec{A}\}+\{\left[\frac{\partial}{\partial t}\,, \vec{\nabla}\right]\varphi\}\times\vec{A}\,,
$$
$$
    \left[\frac{\partial}{\partial t}\,, \vec{\nabla}\cdot\right](\vec{A}\times\vec{B})= \vec{B}\cdot\{\left[\frac{\partial}{\partial t}\,, \vec{\nabla}\times\right]\vec{A}\}-\vec{A}\cdot \{\left[\frac{\partial}{\partial t}\,, \vec{\nabla}\times\right]\vec{B}\}\,.
$$
The differential commutator bracket satisfies the distribution rule
\begin{equation}\label{1}
    \left[\hat{A}\hat{B}\,\,,\hat{C}\right]= \hat{A} \left[\hat{B}\,\,,\hat{C}\right]+ \left[\hat{A}\,\,,\hat{C}\right]\hat{B}\,,
\end{equation}
where
$\hat{A}\,, \hat{B}\,,\hat{C}$ stand for either $\vec{\nabla}$ or $\frac{\partial}{\partial t}$.

It is evident that the differential commutator brackets  identities follow the same ordinary vector identities.
We call the three differential commutator brackets in  Eq.(1) the grad-commutator bracket, the dot-commutator bracket, and the cross-commutator bracket, respectively. The prime idea here is to replace the time derivative of a quantity by the space derivative $\vec{\nabla}$ of another quantity, and vice versa, so that the time derivative of a quantity is followed by a time derivative with which it commutes. We assume here that space and time derivatives don't commute. With this minimal assumption, we have shown here that all physical laws are determined by  vanishing differential commutator bracket.
\section{The continuity equation}
Using quaternionic algebra [4], we have recently found that generalized continuity equations can be written as [5]
\begin{equation}
\vec{\nabla}\cdot \vec{J}+\frac{\partial \rho}{\partial
t}=0\,,
\end{equation}
\begin{equation}
\vec{\nabla}(\rho\,c^2)+\frac{\partial
\vec{J}}{\partial t}=0\,,
\end{equation}
and
\begin{equation}
\vec{\nabla}\times
\vec{J}=0\,,
\end{equation}
where $c$, $\rho$ and $\vec{J}$  are the speed of light, current density, and probability density, respectively.
Now consider the dot-commutator bracket of $\rho\,\vec{J}$
 \begin{equation}\label{1}
    \left[\frac{\partial}{\partial t}\,, \vec{\nabla}\cdot\right](\rho\,\vec{J})=\frac{\partial}{\partial t}\left(\vec{\nabla}\cdot (\rho\,\vec{J})\right)-\vec{\nabla}\cdot\left(\frac{\partial (\rho\,\vec{J})}{\partial t}\right)=0\,.
\end{equation}
Using Eqs.(6), (7), and  (8), and the vector identities
\begin{equation}\label{1}
\vec{\nabla}\cdot(\varphi\vec{G})=(\vec{\nabla}\varphi)\cdot\vec{G}+\varphi(\vec{\nabla}\cdot\vec{G})\,,\qquad
\vec{\nabla}\times(\vec{\nabla}\times\vec{G})=\vec{\nabla}(\vec{\nabla}\cdot\vec{G})-\nabla^2\vec{G}\,
\end{equation}
one obtains
\begin{equation}\label{1}
    \left[\frac{\partial}{\partial t}\,, \vec{\nabla}\cdot\right](\rho\,\vec{J})=c^2\rho\left(\frac{1}{c^2}\frac{\partial^2\rho}{\partial t^2}-\nabla^2\rho\right)+\left(\frac{1}{c^2}\frac{\partial \vec{J}}{\partial t^2}-\nabla^2\vec{J}\right)\cdot\vec{J}=0\,.
\end{equation}
For arbitrary $\rho$ and $\vec{J}$, Eq.(11) yields the two wave equations
\begin{equation}\label{1}
 \frac{1}{c^2}\frac{\partial^2\rho}{\partial t^2}-\nabla^2\rho=0\,,
\end{equation}
and
\begin{equation}\label{1}\frac{1}{c^2}\frac{\partial \vec{J}}{\partial t^2}-\nabla^2\vec{J}=0\,.
\end{equation}
 Equations (12) and (13) are also obtained utilizing the quaternionic formulation following ref. [5]. Hence, the wave equations of $\rho$ and $\vec{J}$, in our present brackets formulation, are equivalent to
\begin{equation}\label{1}
    \left[\frac{\partial}{\partial t}\,, \vec{\nabla}\cdot\right](\rho\,\vec{J})=0\,.
\end{equation}
Equations (12) and (13) show that the charge and current densities satisfy a wave  traveling at the speed of light in vacuum. It is remarkable to know that these two equations are already obtained in ref. [5].
\section{Quantum Mechanics}

Consider a particle described by the four vector $\Psi=(\frac{i}{c}\psi_0\,, \vec{\psi})$. This is equivalent to spinor representation of ordinary quantum mechanics. We have recently developed a quaternionic quantum mechanics dealing with such a four vector [4]. The evolution of this four vector is given by the three equations [4]
\begin{equation}
\vec{\nabla}\cdot\vec{\psi}-\frac{1}{c^2}\frac{\partial \psi_0}{\partial t}-\frac{m_0}{\hbar}\,\psi_0=0\,,
\end{equation}
\begin{equation}
\vec{\nabla}\psi_0-\frac{\partial \vec{\psi}}{\partial t}-\frac{m_0c^2}{\hbar }\,\vec{\psi}=0\,,
\end{equation}
and
\begin{equation}
\vec{\nabla}\times\vec{\psi}=0\,,
\end{equation}
where $m_0$ and $\hbar$ are the quasi-particle mass and Planck constant, respectively.
Equations (15), (16) and (17) yield the two wave equations [5]
\begin{equation}
\frac{1}{c^2}\frac{\partial^2\vec{\psi}}{\partial t^2}-\nabla^2\vec{\psi}+2\left(\frac{m_0}{\hbar}\right)\frac{\partial\vec{\psi}}{\partial t}+\left(\frac{m_0c}{\hbar}\right)^2\vec{\psi}=0\,,
\end{equation}
and
\begin{equation}
\frac{1}{c^2}\frac{\partial^2\psi_0}{\partial t^2}-\nabla^2\psi_0+2\left(\frac{m_0}{\hbar}\right)\frac{\partial\psi_0}{\partial t}+\left(\frac{m_0c}{\hbar}\right)^2\psi_0=0\,.
\end{equation}
 Using the transformation
\begin{equation}
\frac{\partial}{\partial \tau}=\frac{\partial}{\partial t}+\frac{m_0c^2}{\hbar}\,,
 \end{equation}
 so that Eqs.(15) and (16) become
 \begin{equation}
\vec{\nabla}\cdot\vec{\psi}=\frac{1}{c^2}\frac{\partial\psi_0}{\partial \tau}\,,\qquad \vec{\nabla}\psi_0=\frac{\partial\vec{\psi}}{\partial \tau}\,.
 \end{equation}
Employing Eq.(20), Eqs. (18) and (19) are transformed into the wave equations
\begin{equation}
\Box '\,^2\psi_0=0\,,\qquad\,\Box '\,^2\vec{\psi}=0\,\,\,,\qquad{\rm where}\qquad\Box '\,^2=\frac{1}{c^2}\frac{\partial^2}{\partial \tau^2}-\nabla^2\, .
\end{equation}
 Equations (18) and (19) can be obtained from the Einstein's energy equation by setting $E'=E+i\,m_0c^2$, where $E'^2=p^2c^2$ and using the familiar quantum mechanical operator replacements, viz., $\hat{p}=-i\hbar\nabla$ and $\hat{E}=i\hbar\frac{\partial}{\partial t}$. $E'=cp$ is an equation for a massless particle. This is also evident from Eq.(22). Thus, it is interesting that a massive particle can be transformed into a massless particle using Eq.(20). Since energy is a real quantity, this equation is physically acceptable if it describes a particle with imaginary mass. In this case the energy equations split into two parts; one with $E'=E+m_0c^2$ and the other with energy $E'=E-m_0c^2$. Such energies can describe the state of a particle and antiparticle. A hypothetical particle with an imaginary mass moving at a speed higher than the speed of light in vacuum is known as \emph{tachyon} [6]. Hence, our above equation can be used to treat the motion of  tachyons.  This implies that our equations, Eqs.(24) and (27) can be applied to tachyons. Some scientists propose that neutrino can be a tachyonic fermion [7]. We know that the Cherenkov radiation is emitted from a particle moving in a medium with a speed larger than the speed of light in vacuum. When the speed exceeds the speed of light in a vacuum, the extra energy acquired by the particle is transformed in radiation. This can happen momentarily for a particle keeping its total energy conserved. Thus, the excess energy (speed) is such that it compensates the dissipations.

Now consider the cross-commutator bracket of $\psi_0\vec{\psi}$
\begin{equation}\label{1}
\left[\frac{\partial}{\partial t}\,, \vec{\nabla}\times\right](\psi_0\vec{\psi})=\frac{\partial}{\partial t}\left(\vec{\nabla}\times(\psi_0\vec{\psi})\right)-\vec{\nabla}\times\left(\frac{\partial(\psi_0\vec{\psi})}{\partial t}\right)=0\,.
\end{equation}
Using Eqs.(15), (16), and (17), and the vector identities
\begin{equation}\label{1}
\vec{\nabla}\times(\varphi\vec{G})=(\vec{\nabla}\varphi)\times\vec{G}+\varphi(\vec{\nabla}\times\vec{G})\,,\qquad
\vec{\nabla}\times(\vec{\nabla}\times\vec{G})=\vec{\nabla}(\vec{\nabla}\cdot\vec{G})-\nabla^2\vec{G}\,
\end{equation}
yield the wave equation
\begin{equation}
\frac{1}{c^2}\frac{\partial^2\vec{\psi}}{\partial t^2}-\nabla^2\vec{\psi}+2\left(\frac{m_0}{\hbar}\right)\frac{\partial\vec{\psi}}{\partial t}+\left(\frac{m_0c}{\hbar}\right)^2\vec{\psi}=0\,.
\end{equation}
Similarly, the dot-commutator bracket of $\psi_0\vec{\psi}$
\begin{equation}\label{1}
\left[\frac{\partial}{\partial t}\,, \vec{\nabla}\cdot\right](\psi_0\vec{\psi})=\frac{\partial}{\partial t}\left(\vec{\nabla}\cdot(\psi_0\vec{\psi})\right)-\vec{\nabla}\cdot\left(\frac{\partial(\psi_0\vec{\psi})}{\partial t}\right)=0\,.
\end{equation}
Upon using Eqs.(10), (15), and (16), one obtains the wave equation of $\psi_0$
\begin{equation}
\frac{1}{c^2}\frac{\partial^2\psi_0}{\partial t^2}-\nabla^2\psi_0+2\left(\frac{m_0}{\hbar}\right)\frac{\partial\psi_0}{\partial t}+\left(\frac{m_0c}{\hbar}\right)^2\psi_0=0\,.
\end{equation}
It is interesting to see that Eqs. (25), and (27) are the same as Eqs.(18), and (19) obtained from quaternionic manipulation. We thus write Eqs.(25), and (27) as
\begin{equation}\label{1}
\left[\frac{\partial}{\partial t}\,, \vec{\nabla}\times\right](\psi_0\vec{\psi})=0\,\,,\qquad \left[\frac{\partial}{\partial t}\,, \vec{\nabla}\cdot\right](\psi_0\vec{\psi})=0\,.
\end{equation}
\section{Dirac's equation}
Dirac's equation can be written in the form [1]
\begin{equation}\label{1}
\frac{1}{c}\frac{\partial\psi}{\partial t}+\vec{\alpha}\cdot\vec{\nabla}\psi+\frac{im_0c\,\beta}{\hbar}\psi=0\,.
\end{equation}
Consider the differential commutator bracket
\begin{equation}\label{1}
\left[\frac{\partial}{\partial t}\,\,,\, \vec{\alpha}\cdot\vec{\nabla}\right]\psi=\frac{\partial}{\partial t} (\vec{\alpha}\cdot\vec{\nabla}\psi)-(\vec{\alpha}\cdot\vec{\nabla})\frac{\partial \psi}{\partial t}=0   \,.
\end{equation}
Using Eq.(29), Eq.(30) yields
\begin{equation}
\frac{1}{c^2}\frac{\partial^2\psi}{\partial t^2}-\nabla^2\psi+2\left(\frac{m_0i\beta}{\hbar}\right)\frac{\partial\psi}{\partial t}-\left(\frac{m_0c}{\hbar}\right)^2\psi=0\,,
\end{equation}
where we have used the fact that  $\beta=\left (\begin{array}{cc}
  1 & 0 \\
  0 & -1 \\
\end{array}\right),$    $\alpha=\left (\begin{array}{cc}
 0 & \vec{\sigma}   \\
   \vec{\sigma}  & 0 \\
\end{array}\right),$ $\alpha^2=\beta^2=1$ and $\vec{\sigma}$ are the Pauli matrices. Equation (31) can be obtained from Eq.(29) by squaring it. This equation can be compared with the Klein-Gordon equation of spin - 0 particles
$$
\frac{1}{c^2}\frac{\partial^2\psi}{\partial t^2}-\nabla^2\psi+\left(\frac{m_0c}{\hbar}\right)^2\psi=0\,.
$$
Equation (31) is another form of Dirac's equation exhibiting the wave nature of spin-$\frac{1}{2}$ particles explicitly.
Using the transformation
\begin{equation}
\frac{\partial}{\partial \eta}=\frac{\partial}{\partial t}+i\frac{m_0c^2}{\hbar}\beta\,,
\end{equation}
Eq.(31) can be written as
\begin{equation}
\frac{1}{c^2}\frac{\partial^2\psi}{\partial \eta^2}-\nabla^2\psi=0\,.
\end{equation}
This is a wave equation for a massless particle. Thus, a particle annihilates (loses its mass) after a time interval of $\Delta t=\frac{\hbar}{m_0c^2}$ and then created (acquired a mass). It is interesting to notice that during such a period of time, energy can be violated as endorsed by the Heisenberg's uncertainty relation ($\Delta t\,\Delta E\ge \hbar$) where $\Delta E=2m_0c^2$. This also applies to the particles as defined by Eq.(20). Equation (31) describes the behavior of a particle of a definite mass $m_0$. After a time of $\Delta t=\frac{\hbar}{m_0c^2}$ the particle becomes a wave with energy $E'=pc$ governed by Eq.(33). The particle interacts with the vacuum in such a way that when the particle becomes a wave (annihilates) gives its mass energy to the vacuum, and restores it after a time of $\Delta t$ as defined before becoming a particle once again. This is the essence of the oscillatory motion as known as \emph{zitterbewegung} motion [1]. This result supports the fact that there is a vacuum fluctuation associated with the particle. This means when a particle becomes a wave it gives its mass to the vacuum and restores it when becomes a corpuscule. Thus, the corpuscular and wave nature (duality) of a particle is concomitant with the particle motion.
Since $\psi$ is a four components spinor, we can write it in terms of two components doublets, viz., $\psi=\left(\begin{array}{c}
\psi_+ \\
\psi_- \\
\end{array}
\right)$.
Substituting these decomposed spinors in Eq.(31), one obtains
the two equations
\begin{equation}
\frac{1}{c^2}\frac{\partial^2\psi_+}{\partial t^2}-\nabla^2\psi_++2\left(\frac{m_0i}{\hbar}\right)\frac{\partial\psi_+}{\partial t}-\left(\frac{m_0c}{\hbar}\right)^2\psi_+=0\,,
\end{equation}
and
\begin{equation}
\frac{1}{c^2}\frac{\partial^2\psi_-}{\partial t^2}-\nabla^2\psi_--2\left(\frac{m_0i}{\hbar}\right)\frac{\partial\psi_-}{\partial t}-\left(\frac{m_0c}{\hbar}\right)^2\psi_-=0\,,
\end{equation}
Equation (34) and (35) imply two energy solutions, one with $E'=E-m_0c^2$ and the other with energy $E'=E+m_0c^2$. This is also evident from using the Einstein energy-momentum equation $(E^2=p^2c^2+m_0^2c^4$). The two energy states  may define a particle and an antiparticle. Since the time factor in the wavefunction is of the form $\exp(-iEt/\hbar)$, the new wavefunction with the new time ($\eta$) will become $\exp(-iE'\eta/\hbar)$, where
 \begin{equation}
\eta=t-i\frac{m_0c^2}{\hbar}\,,
\end{equation}
is a complex time, as evident from Eq.(33). It can be seen as a rotation of the real time by a phase into a complex plane.
 Such an effect arises from the very nature of the particle when  propagating in space-time. The third term in Eq.(31) represents a dissipation that may result from the motion of the particle in space (ether). Hence, any massive particle should exhibit this sort of propagation when travels in space-time. This term is vanishingly small compared with the mass term in Eq.(31) but very fundamental. Moreover, Our equations (25) and (27) are equivalent to Dirac equations, Eqs.(34) and (35), if we replace $m_0$ by $-im_0$.

 Consider now the case when $\psi$ is space independent so that Eq.(31) becomes
 \begin{equation}
\frac{d^2\psi}{d t^2}+2\left(\frac{m_0c^2i\beta}{\hbar}\right)\frac{d\psi}{d t}-\left(\frac{m_0c^2}{\hbar}\right)^2\psi=0\,.
\end{equation}
with $\psi=\left(  \begin{array}{c}
\psi_+ \\
\psi_- \\
\end{array}
\right)$ this yields the two equations
\begin{equation}
\frac{d^2\psi_+}{d t^2}+2\left(\frac{m_0c^2i}{\hbar}\right)\frac{d\psi_+}{d t}-\left(\frac{m_0c^2}{\hbar}\right)^2\psi_+=0\,,
\end{equation}
and
\begin{equation}
\frac{d^2\psi_-}{d t^2}-2\left(\frac{m_0c^2i}{\hbar}\right)\frac{d\psi_-}{d t}-\left(\frac{m_0c^2}{\hbar}\right)^2\psi_-=0\,.
\end{equation}
These two equations have an oscillatory behavior, i.e.,
\begin{equation}
\psi_+(t)=A_+\exp(-i\omega\,t)\,\,,\qquad \psi_-(t)=A_-\exp(i\omega\,t)\,\qquad A_+\,\,, A_-=\rm const.\,,
\end{equation}
where $\omega=\frac{m_0\,c^2}{\hbar}$. This means that the particle with the wavefunction $\psi$ has two energy eigen states, one for a particle and the other one for an antiparticle. Hence, Eq.(40) reveals that the particle is described by a standing wave having positive and negative energy. This is the essence of Dirac's theory. The two states are separated by an amount of energy, $\Delta E'=2m_0c^2$.

Using Eq.(29), Eq.(31) can be written in the form
\begin{equation}
\frac{1}{c^2}\frac{\partial^2\psi}{\partial t^2}-\nabla^2\psi-2\left(\frac{m_0c\,i}{\hbar}\right)\beta\vec{\alpha}\cdot\vec{\nabla}\psi+\left(\frac{m_0c}{\hbar}\right)^2\psi=0\,,
\end{equation}
This can be written as
 \begin{equation}
\frac{1}{c^2}\frac{\partial^2\psi}{\partial t^2}-\nabla'\,^2\psi=0\,,
\end{equation}
where
 \begin{equation}
\vec{\nabla}'=\vec{\nabla}+i\frac{m_0c\beta}{\hbar}\,\vec{\alpha}\,.
\end{equation}
 Equation (42) is a wave equation in the new coordinate defined by  Eq.(43). Equation (43) can be written as
\begin{equation}
\vec{p}'=\vec{p}+m_0c\beta\vec{\alpha}\,\,.
\end{equation}
 This can be compared with the covariant derivative that results from the interaction of a particle with a photon field $\vec{A}$, viz., $\vec{p}'=\vec{p}-e\vec{A}$. Equation (32) can be written as
 \begin{equation}
E'=E-m_0c^2\beta\,\,.
\end{equation}
Equations (32) and (43) can be combined into a single equation as
\begin{equation}
D_\mu=\partial_\mu+i\frac{m_0c}{\hbar}\,\gamma_\mu\,,\qquad \gamma_\mu=(\beta\,,\beta\alpha)\,.
\end{equation}
We call here the derivative $D'_\mu$ the spinor derivative. With this derivative the Dirac equation takes the simple  forms
\begin{equation}
\slash\!\!\!\!P\,\psi=0\,,\qquad{\rm where}\qquad \slash\!\!\!\!P_\mu=i\hbar\,\gamma^\mu D_\mu\,,
\end{equation}
It is interesting to notice that Eqs.(47) looks like  massless Dirac equation.
The second term in Eq.(44) represents a self interaction of the particle due to its spin. Since the vector potential is a gauge field, the spin of the particle will accordingly becomes a gauge quantity. In present  case, the electron interacts with its spin that is related to  $\vec{\alpha}$. This effect represents a self interaction of the particle. The algebra of the $P_\mu$'s commutator bracket is
\begin{equation}
[P^\mu\,, P^\nu]=2m_0c\,i\,\sigma^{\mu\nu}\,,\qquad \sigma^{\mu\nu}=\frac{i}{2}\,[\gamma^\mu\,\,,\gamma^\nu]\,.
\end{equation}
Thus, unlike partial derivative, spinor derivatives do not commute. The momenta commute for a massless particle.
Equation (41) describes a particle with definite mass which after a characteristic distance of $\ell=\frac{\hbar}{m_0c}$ becomes a wave as described by Eq.(42). Hence, the corpuscular  nature of the particle is exhibited after a distance of $X_0=\frac{2\hbar}{m_0c}$, and the wave nature after a time of $T_0=\frac{2\hbar}{m_0c^2}$. The particle's velocity must be in such a way to reach the next point in the same time required to be in the other state. This requires its velocity to be $\vec{v}=\vec{\alpha}\, c$. Thus, the particle remains in a continuous dual state (particle+wave). This duality is manifested during a time of $T_0$ at a distance of $X_0$. This may usher into a quantization of space and time in units of  $T_0$ and $X_0$ as fundamental units. With the definition $\vec{v}=\vec{\alpha}\, c$, Dirac's equation can be written as
\begin{equation}
\frac{\partial}{\partial t}+\vec{v}\cdot\vec{\nabla}=-\frac{im_0c^2}{\hbar}\beta=\frac{d}{dt}\,
\end{equation}
which implies that
\begin{equation}
 i\hbar\frac{d\psi}{dt}=m_0c^2\beta\,\psi\,.
 \end{equation}
 Hence, Equation (50) is a variant form of Dirac's equation.
But since $\psi$ can be written as two-components column, viz., $\psi=\left(\begin{array}{c}
\psi_+ \\
\psi_-
\end{array}\right)$, the above equation implies that
\begin{equation} i\hbar\frac{d\psi_+}{dt}=m_0c^2\,\psi_+\,,\qquad i\hbar\frac{d\psi_-}{dt}=-m_0c^2\psi_-\,.\end{equation}
 This shows that the operator $\hat{A}=i\hbar\frac{d}{dt}$ is the rest mass energy operator of the individual spinor components.
\subsection{The continuity equation}
Taking the complex conjugate of Eq.(31) and multiplying it by $\psi$ once  from right, and subtract it from Eq.(31) after multiplying it by $\psi^*$ from left, we obtain the continuity equation
 \begin{equation}
\frac{\partial \rho_{_T}}{\partial t}+\vec{\nabla}\cdot\vec{J}=0\,,
\end{equation}
where
\begin{equation}
\rho_{_T}=\rho_{_S}+\rho_{_{KG}}\,\,,\, \rho_{_S}=\beta\psi^*\psi\,,\, \rho_{_{KG}}=\frac{i\,\hbar}{2m_0c^2}\left(\psi^*\frac{\partial\psi}{\partial t}-\psi\frac{\partial\psi^*}{\partial t}\right), \vec{J}=\frac{\hbar}{2m_0i}\left(\psi^*\vec{\nabla}\psi-\psi\vec{\nabla}\psi^*\right)\,.
\end{equation}
It is understood here that $\psi$ is a spinor, $\rho_{_T}$ is the charge density and $\vec{J}$ is the current density. It is interesting that Eq.(31), obtained from Dirac's equation using the differential operator bracket in Eq.(30), yields a continuity equation sharing both the Dirac and Klein-Gordon features of the charge (probability) density. This interplay exists despite the fact that Dirac's equation represents a fermionic particle while Klein-Gordon equation represents a bosonic particle.
\section{The space and time  invariance of Dirac's equation}
If we apply the transformations in Eq.(32) and (43) to Eq.(29), Dirac's equation will be invariant. Thus, the space and time transformation represented by Eqs.(32) and (43) ushers into  a new transformation of Dirac's equation that were never known before. With some scrutiny, we know from the theory of relativity that the kinetic energy $(E_K)$ is related to the total energy $(E$) by $E_K=E-m_0c^2$. In quantum mechanics, $E\rightarrow i\hbar\frac{\partial}{\partial t}$, so that
\begin{equation}
\hat{E}_K=i\hbar\frac{\partial}{\partial t}-m_0c^2=i\hbar\left(\frac{\partial}{\partial t}+i\frac{m_0c^2}{\hbar}\right)=i\hbar\frac{\partial}{\partial \eta}\,.
\end{equation}
This is the relativistic kinetic energy operator. Alternatively, using Eq.(29), this can be written as
\begin{equation}
\hat{E}_K=c\vec{\alpha}\cdot\vec{\hat{p}}\,.
\end{equation}
This equation implies that Dirac's equation can be obtained from the relativistic energy equation
\begin{equation}
E=E_K+\beta\,m_0c^2\,.
\end{equation}
This equation suggests that there are two possible energy equations. These are $E_\pm=E_K\pm \,m_0c^2$. Hence, a Dirac's particle has in principle two energies, $E_+=E_K+m_0c^2$ and $E_-=E_K-m_0c^2$.

Using Eq.(32), Dirac's equation Eq.(29), becomes
\begin{equation}
\frac{\partial\psi}{\partial \eta}+\vec{\nabla}\cdot (c\vec{\alpha})\psi=0\,.
\end{equation}
Thus, in the time coordinate $\eta$, Dirac's equation represents a continuity-like equation.
However, in the real time, the continuity equation in Dirac's formalism, is defined as
\begin{equation}
\frac{\partial\rho}{\partial t}+\vec{\nabla}\cdot \vec{J}=0\,,
\end{equation}
where $\rho=\psi^+\psi$ and $\vec{J}=\psi^+c\vec{\alpha}\psi$. Using Eq.(43), Dirac's equation is transformed into a continuity-like equation in the new space coordinate, viz.,
\begin{equation}
\frac{\partial\psi}{\partial t}+\vec{\nabla}'\cdot (c\vec{\alpha})\psi=0\,.
\end{equation}
Notice here that $\vec{J}$ has the same form in both coordinates.
\section{Space and time invariance of Maxwell's equations}
We would like here to apply the space and time transformations in Eqs.(32) and (43) to explore their implications in Maxwell's equations.
These transformations leave Dirac's equation invariant. We know that quantum electrodynamics incorporates the interaction of an electron  with a photon. Quantum electrodynamics becomes invariant under gauge transformation, if we replace the partial derivative  with a covariant derivative incorporating the photon field. Analogously, we assume here that Maxwell's equations are invariant under the new space and time transformations in Eqs.(32) and (43). Applying Eq.(32) and (43) to the Ampere's and Faraday's equations [8]
\begin{equation}
\vec{\nabla}\times\vec{E}=-\frac{\partial \vec{B}}{\partial t}\,,
\end{equation}
and
\begin{equation}
\vec{\nabla}\times\vec{B}=\mu_0\vec{J}+\frac{1}{c^2}\frac{\partial \vec{E}}{\partial t}\,,
\end{equation}
yield
\begin{equation}
\vec{B}=\frac{\vec{v}\times\vec{E}}{c^2}\,,\qquad \vec{v}=c\vec{\alpha}\,,
\end{equation}
and
\begin{equation}
\vec{E}=-\vec{v}\times\vec{B}\,,\qquad \vec{v}=c\vec{\alpha}\,.
\end{equation}
The remaining two Maxwell's equations
\begin{equation}
\vec{\nabla}\cdot\vec{E}=\frac{\rho}{\epsilon_0}\,,\qquad \vec{\nabla}\cdot\vec{B}=0\,.
\end{equation}
yield
\begin{equation}
\vec{v}\cdot\vec{E}=0\,,\qquad \vec{v}\cdot\vec{B}=0\,,\qquad \vec{v}=c\vec{\alpha}.
\end{equation}
It is interesting to notice that  Eq.(65) is compatible with Eqs.(62) and (63). Moreover, Eqs.(62) and (63) define the relations between the electric and magnetic fields produced by the moving charge. If the electric (magnetic) field is known, one can obtain the corresponding magnetic (electric) field. Equation (63) shows that the charge moving with constant velocity experiences no net force.
The electric field lines of a moving charge crowded in the direction perpendicular to $\vec{v}$ and are given by [9]
\begin{equation}
\vec{E}=\frac{1}{4\pi\varepsilon_0}\frac{q(1-\beta^2)}{(1-\beta^2\sin^2\theta)^{\frac{3}{2}}}\frac{\vec{r}}{r^3}\,,\qquad \beta=\frac{v}{c}\,,
\end{equation}
and
\begin{equation}
\vec{B}=\frac{\mu_0}{4\pi}\frac{q(1-\beta^2)}{(1-\beta^2\sin^2\theta)^{\frac{3}{2}}}\frac{\vec{v}\times\vec{r}}{r^3}\,,\qquad \beta=\frac{v}{c}.
\end{equation}
Equation (66) shows that the electric and magnetic fields of a moving charge in the forward direction ($\theta=0)$ are less than the electric field of stationary charge. However, the electric and magnetic fields in the perpendicular direction ($\theta=\pi/2)$ are bigger than the electric field of  stationary charge.
Equations (66) and (67) give the relations between the   electric and magnetic fields produced by a moving charged particle with constant velocity $\vec{v}$. This velocity is given by $\vec{v}=c\, \vec{\alpha}$. This coincides with the quantum mechanics definition of the particle velocity [1]. Equation (65) shows that the electric and magnetic fields produced by the charged particle are always perpendicular to the particle's direction of motion.  Equation (67) gives, at low velocity,  the Biot-Savart law.  The power delivered by the fermionic charged particle by its  electric and magnetic fields is given by $P=\vec{F}\cdot\vec{v}=q\vec{v}\cdot\vec{E}+q\vec{v}\cdot(\vec{v}\times\vec{B})=0$.
The application of the transformations (32) and (43) in the generalized continuity equations (6) and (8) yields
\begin{equation}
\vec{v}\cdot\,\vec{J}=-\rho\, c^2\,,\qquad \vec{J}=-\,\rho\,\vec{v}\,,\qquad \vec{v}=c\vec{\alpha}\,.
\end{equation}
Since in Dirac formalism $\rho>0$, the current ushers in a direction opposite to the velocity direction. Moreover, for a constant velocity, one has $(\vec{\nabla}\rho)\times\vec{v}=0$. Equation (68) is very interesting since it defines the charge density (scalar) in terms of the current density (vector). Accordingly, one can define the four vectors in terms of vectorial quantities only.
\section{Concluding Remarks}
By introducing  three vanishing differential commutator brackets for spinor fields,  we have derived  a variant form of Dirac's and  Klein-Gordon wave equations. Dirac's equation yields a modified Klein-Gordon wave equation. This equation yields directly two energy states for the particle in question. Moreover, Dirac's equation is found to be similar to the continuity equation. We have found time and space transformations under which Dirac's and Maxwell's equations are invariant. In terms of these coordinates, Dirac and Klein-Gordon equations describe a massless particle. The invariance of these transformations under Maxwell's equations  shows that the electric and magnetic fields produced by a moving charge are perpendicular to velocity of the particle. Hence, there is no power associated with these fields. The space and time transformations show that space and time are quantized in terms of characteristic units of\,  $T_0$ and $X_0$. The fermionic charged particle exhibits its wave and corpuscular nature on periodic space and time basis.
\section*{Acknowledgements}

\vspace*{-2pt}
This work is supported by the university of Khartoum research fund. We gratefully acknowledge this  support. The critical and useful comments by the anonymous referees are highly acknowledged.

\section*{References}
$[1]$ Bjorken, J. D., and Drell, S. D., \emph{Relativistic Quantum Mechanics}, McGraw-Hill, (1964);
 Landau, L.D. and  Liftshiz, E.M., \emph{Quantum Mechanics}, 3rd ed. Pergamon Press, Oxford, 1977; Berestetskii, V. B., Pitaevskii, L. P., and  Lifshitz, E.M.,  \emph{Quantum Electrodynamics}, 2nd. ed. Vol.4, 1982.\\
$[2]$ Arbab, A. I., and Satti, Z., \emph{Progress in Physics}, \textbf{2}, 8 (2009).\\
$[3]$ Arbab, A. I., and Yassein, F. A., \emph{A New Formulation of Electromagnetism}, http://arxiv.org/abs/1002.4706  (to appear in \emph{Journal of Electromagnetic Analysis and Applications}, 2010).\\
$[4]$ Arbab, A. I.,  \emph{A Quaternionic Quantum Mechanics}, http://arxiv.org/abs/1003.0075;  Tait, P. G., \emph{An elementary treatise on quaternions}, 2nd ed., Cambridge University Press (1873);  Harmuth, H. F., Barrett, T. W., and Meffert, B., \emph{Modified Maxwell equations in quantum electrodynamics, Singapore}; River Edge, N.J., World Scientific, (2001).\\
$[5]$ Arbab, A. I., and Widatallah, H. M., \emph{The generalized continuity equations}, Chin. Phys. Lett. Vol. 27, No. 8 (2010); DOI: 10.1088/0256-307X/27/8/084703.\\
$[6]$ Feinberg, G., \emph{Possibility of Faster-Than-Light Particles}, Phys. Rev. \textbf{159}, 1089 (1967).\\
$[7]$ Ciborowski, J., \emph{Hypothesis of tachyonic neutrinos}, Acta Physicsa Polonica, \textbf{B.29}, 113 (1998).\\
$[8]$ Jackson, J, D., \emph{Classical Electrodynamics}, Wiley, New York, 2nd edition, (1975).\\
$[9]$ Zbigniew, F., \emph{Lecture notes in electromagnetic theory}, University of Queensland, ( 2005).
\newpage
\section*{Appendix}
$$
    \left[\frac{\partial}{\partial t}\,, \vec{\nabla}\right]\varphi=\frac{\partial}{\partial t}\left(\vec{\nabla}\varphi\right)-\vec{\nabla}\left(\frac{\partial \varphi}{\partial t}\right)\,\qquad\qquad\qquad\qquad \qquad(A1)
$$
$$
    \left[\frac{\partial}{\partial t}\,, \vec{\nabla}\cdot\right]\vec{F}=\frac{\partial}{\partial t}\left(\vec{\nabla}\cdot\vec{F}\right)-\vec{\nabla}\cdot\left(\frac{\partial \vec{F}}{\partial t}\right)\,\qquad \qquad\qquad\qquad (A2)
$$
$$
  \left[\frac{\partial}{\partial t}\,, \vec{\nabla}\times\right]\vec{F}=\frac{\partial}{\partial t}\left(\vec{\nabla}\times\vec{F}\right)-\vec{\nabla}\times\left(\frac{\partial \vec{F}}{\partial t}\right)\,\qquad\qquad\qquad\qquad (A3)
$$

$$
    \left[\frac{\partial}{\partial t}\,\,,\vec{\nabla}\times\right](\vec{F}\times\vec{G})=\vec{F}\times\left(\frac{\partial}{\partial t}(\vec{\nabla}\times\vec{G})-\vec{\nabla}\times\frac{\partial\vec{G}}{\partial t}\right)+\left(\vec{\nabla}\times\frac{\partial\vec{F}}{\partial t}-\frac{\partial}{\partial t}(\vec{\nabla}\times\vec{F})\right)\times\vec{G}\qquad (A4)
$$
$$
    \left[\frac{\partial}{\partial t}\,\,,\vec{\nabla}\right](\vec{F}\cdot\vec{G})=\vec{F}\left(\frac{\partial}{\partial t}(\vec{\nabla}\cdot\vec{G})-\vec{\nabla}\cdot\frac{\partial\vec{G}}{\partial t}\right)+\left(\vec{\nabla}\cdot\frac{\partial\vec{F}}{\partial t}-\frac{\partial}{\partial t}(\vec{\nabla}\cdot\vec{F})\right)\vec{G}\qquad\qquad\qquad\qquad (A5)
$$
$$
\left[\frac{\partial^2}{\partial t^2}\,, \vec{\nabla}\right]\vec{\varphi}= \left[\frac{\partial}{\partial t}\,, \vec{\nabla}\right]\frac{\partial \varphi}{\partial t}\,, \left[\frac{\partial^2}{\partial t^2}\,, \vec{\nabla}\cdot\right]\vec{F}= \left[\frac{\partial}{\partial t}\,, \vec{\nabla\cdot}\right]\frac{\partial \vec{F}}{\partial t}\,,  \left[\frac{\partial^2}{\partial t^2}\,, \vec{\nabla}\times\right]\vec{F}= \left[\frac{\partial}{\partial t}\,, \vec{\nabla\times}\right]\frac{\partial \vec{F}}{\partial t}\,(A6)
$$
\end{document}